\documentstyle[preprint,aps]{revtex} 
\tightenlines
\include{epsf} 
\begin{document} 
%
\def\Journal#1#2#3#4{{#1} {\bf #2}, #3 (#4)} 
 
\def\TAJ{\em The Astrophysical Journal} 
\def\NCA{\em Nuovo Cimento} 
\def\NIM{\em Nucl. Instrum. Methods} 
\def\NIMA{{\em Nucl. Instrum. Methods} A} 
\def\NPB{{\em Nucl. Phys.} B} 
\def\NPA{{\em Nucl. Phys.} A} 
\def\NP{{\em Nucl. Phys.} } 
\def\PLB{{\em Phys. Lett.} B} 
\def\PRL{\em Phys. Rev. Lett.} 
\def\PRD{{\em Phys. Rev.} D} 
\def\PRC{{\em Phys. Rev.} C} 
\def\PRA{{\em Phys. Rev.} A} 
\def\PR{{\em Phys. Rev.} } 
\def\PRep{{\em Phys. Rep.}} 
\def\ZPC{{\em Z. Phys.} C} 
\def\SJP{{\em Sov. Phys. JETP}} 
\def\SJNP{{\em Sov. Phys. Nucl. Phys.}} 
\def\FBS{{\em Few Body Systems Suppl.}} 
\def\IJMP{{\em Int. J. Mod. Phys.} A} 
\def\UJP{{\em Ukr. J. of Phys.}} 
\def\CJP{{\em Can. J. Phys.}} 
\def\SCI{{\em Science} } 
\def\APJ{{\em Ap. J.}} 
\title{Bremsstrahlung of Flavor-Degenerate Pairs by\\Neutrinos in the 
Nuclear Field} 
\author{L. Chatterjee$^{1,2}$, M. R. Strayer$^2$, and J. S. Wu$^{3,2}$} 
\address{$^1$ Physics Department, Cumberland University, Lebanon, TN 37087\\ 
$^2$Physics Division, Oak Ridge National Laboratory, Oak Ridge, TN 37831-6373\\ 
$^3$Dept.\ of Natural Sciences, Fayetteville State University, Fayetteville, NC 
28301-4298} 
\maketitle 

\begin{abstract} 
 
Neutrino bremsstrahlung of flavor-degenerate pairs in the field of a nucleus 
is of potential importance for neutrino astrophysics and is representative of a class of processes connecting leptonic electroweak sectors to real or virtual photons. We focus on first generation flavor production by both electron and muon neutrinos and present Standard Model cross sections and distributions for lead and iron nuclei. The results  
(of order 10$^{-41}$ cm$^2$ for $\nu_{e}$ - 
lead collisions at 100 MeV) have been fitted to empirical formulae that can be used to estimate backgrounds to neutrino detection experiments and flux normalizations. A compact form of the matrix element obtained by analytic reduction is used to explain the distributions. The V-A limits of the cross sections are shown to agree with published work from the pre-neutral current era. 
 Event signatures and the possible roles of these  
processes in stellar and laboratory neutrino physics are discussed. Cross sections are compared with those for neutrino-electron scattering for neutrino spectra corresponding to typical supernovae temperatures.  
 
\end{abstract} 

\section{Introduction} 
 
       The importance of neutrinos in our fundamental understanding of the physical universe is well recognized.  Neutrinos, as neutral leptons, 
interact with matter only via the W and Z members of the electroweak gauge boson family. The intrinsic properties of the neutrino play a critical role in the evolution and dynamics of the universe and neutrino reactions influence the life cycles of stars from infancy to possibly explosive death. Through their dynamic role in stellar astrophysics and cosmology, neutrinos also affect the production and distribution of elements and serve as valuable informants on the cosmos and its stellar inhabitants. As elementary particles and laboratory probes, neutrinos can unravel nature and matter at various levels of structure from atoms and nuclei to valence and sea quarks. 
 
        The established importance of neutrino-nucleus and neutrino-lepton electroweak processes in stellar and laboratory physics has motivated increased interest in achieving higher accuracy in the theoretical and experimental determination of these rates. This interest has 
stimulated exploration of the possible effects of related neutrino induced 
electroweak channels that may be competitive to the leading channels or 
affect their observation and detection
\cite{kltv,avi,lsw,schin,dic72,dic97,bern,lee,bahc95,sche83,kl,hax,woo90}. 

Bremsstrahlung of charged lepton-antilepton pairs by neutrinos in the electromagnetic field of nuclei is representative of a class of processes, which produce lepton-antilepton pairs in the electroweak sector and couple these electromagnetically to nuclei. As the nucleus participates via virtual photon exchange, the resultant cross sections are of order (GZ$\alpha$)$^2$, where G is the Fermi constant and (Z$\alpha$) represents the electromagnetic coupling to the  
nucleus. These may be compared to processes of order (G$^{2}\alpha$), 
including   
neutrino-photon collisions \cite{schin,dic72,dic97}, and alpha order corrections to the lowest order electroweak channels \cite{bahc95,bern,lee}. The contributions from pair  
production could therefore surpass the (G$^{2}\alpha$) ones for values of Z 
exceeding 12, depending on the other effects such as phase  
space. They are also comparable with neutrino-electron scattering channels 
\cite{sche83} at sufficiently high Z values and could be manifested as backgrounds to neutrino cross section and flux measurements as well as neutrino detection experiments. The leading order neutrino-nucleus  
cross sections, behaving as (G$^{2}$A)\cite{kl,hax,woo90,con72} yield larger cross sections than those from  pair-production channels. 
 
The neutrino bremsstrahlung in the nuclear field can produce flavor-asymmetric or symmetric pairs of all flavors, depending on the incident neutrino energy and flavor.  
              Restricting exit channels to flavor degenerate charged pairs and including all flavor varieties for the incident neutrinos, the allowed channels 
are:
\begin{eqnarray} 
\nu_{e} + A & \rightarrow & \nu_{e}^{'} + A + l^{+} + l^{-},  \nonumber \\
\nu_{\mu} + A & \rightarrow & \nu_{\mu}^{'} + A + l^{+} + l^{-}, \nonumber  \\
\nu_{\tau} + A & \rightarrow & \nu_{\tau}^{'} + A + l^{+} + l^{-}, 
\end{eqnarray} 
where $l^{+}$ and $l^{-}$ refer to the produced charged lepton-antilepton pair,
and  A to the nucleus. 

In this paper, we focus on the production of first-generation flavor pairs only and consider incident neutrinos of electron and muon  
flavor. The tau neutrino rates would be the same as those for the muon neutrino. The formalism is in the framework of the Standard Model and valid for all combinations of flavor-degenerate and non-degenerate pairs  
with appropriate choice of parameters. Due to the lower threshold, flavor-degenerate first-generation pairs are those expected to be of  
importance for stellar astrophysics or low-energy laboratory neutrino physics. The electron neutrinos produce electron--positron pairs by both charged and neutral current channels, while the muon  
neutrino produces these pairs only via neutral current interactions.
 We restrict our study to spin-zero nuclei that respond coherently and elastically.  
 
Processes of this type have been calculated in the pre-neutral current era \cite{Cyz,Koz,Sha,Mar,okun,angle}, but the literature displays small disagreements between the various  
cross sections. In light of the topical importance of stellar and laboratory neutrino physics, we were motivated to provide a correct description  
of these processes, including the neutral current channels, and assess their astrophysical and laboratory importance. We have computed the  
cross sections for these processes in the Standard Model as a function of incident neutrino energy and compare the V-A limit with earlier  
work \cite{Cyz}, as the earlier work is restricted to this approximation.
We have fitted the numerical Standard Model cross sections for production of electron-positron pairs by electron and muon neutrinos to determine empirical formulae that can be used for various applications.  
Angular distributions for all final state leptons and energy distributions for the produced pair are also reported.
The coupled electroweak and photon vertices involved in the processes under investigation are also of interest from the point of view of  
fundamental physics and the parameters of the electroweak theory. We use analytic approximations to the cross sections in order to understand these  
physics issues and explain the energy and angular distributions. Possible roles of the reaction in stellar and laboratory neutrino physics are discussed. The results can be used to estimate effects of the process on neutrino science. 
The remainder of this paper is divided into the following sections: 
Bremsstrahlung of electron flavor pairs in neutrino-nucleus collisions; the formalism; Standard Model cross sections; analytic reduction; distributions;  experimental  
signatures and possible roles in neutrino science. 
 
\section{Bremsstrahlung of electron flavor pairs in neutrino-nucleus collisions} Electron-flavor pair production in neutrino-nucleus collisions can  
proceed by both charged and neutral electroweak currents. The incident neutrino flavors allowed by the Standard Model are necessarily  
different for the two types of currents, and the Feynman diagrams for such processes include a purely leptonic electroweak sector coupled to  
the  nucleus via a virtual photon. 
The channels can be represented by the general equation Eq. (1).
The Feynman diagram for charged current mediated electron-positron pair bremsstrahlung is shown in Fig.~1 along with the cross term for  
electron neutrino-initiated channels. The neutral current processes allow flavor freedom to the incident neutrino and can be described by the  
Feynman diagram of Fig.~2. 
It may be noted that the electroweak sectors in Figs.~1 and ~2
 represent purely leptonic currents interacting via the electroweak gauge  
bosons. In the absence of the nuclear legs and their intermediary virtual photon, the electroweak sector would describe the  
process, $\nu \rightarrow \nu$ + e$^{-}$ + e$^{+}$, analogous to muon decay. Such a channel would satisfy Standard Model charge and lepton-number  
conservation by a suitable choice of the participating lepton flavors and charges and could be compared with other purely leptonic four-fermion electroweak channels. However, it cannot be realized physically due to kinematic  
and momentum constraints. 
 
The presence of the virtual photon and nuclear sector make the pair production process on-shell, and its interference with the electroweak  
sector determines the correlations and physics of these processes. 
We demonstrate the presence and survival of the individual sectors via  
an analytic reduction of the matrix element. This also permits analysis of the distributions in the light of helicity constraints similar to those appearing in purely leptonic electroweak modes. The cutting of the nuclear legs in 
Figs.~1 and ~2  transforms the virtual photon into a real one, and  
the diagrams then correspond to similar pair production channels in neutrino-photon collisions, which can be described by an analogous  
formalism. We discuss these issues further in a later section. 
In contrast to the processes described by Figs.~1 and ~2 
and Eq.\ (1), the neutrino-nucleus reactions believed to dominate explosive star  
death are the lowest order electroweak processes corresponding to direct exchange of W or Z gauge bosons between the neutrino and the  
nucleus and contributing to order (G$^{2}$A) 
\cite{kl,hax,woo90,con72}. 
\section{The Formalism} 
\subsection{The Electroweak Sector and the Effective Lagrangian} Neutrinos produced in supernova environments, solar interiors, and those  
involved in laboratory experiments have relatively low energies, and this restricts the four-momentum transferred at the electroweak  
sector to values well below the masses of the electroweak gauge bosons. Under these circumstances, the gauge bosons may be collapsed  
to point coupling. As we are primarily interested in such neutrinos, we will carry out the numerical calculations in the effective Lagrangian  
approximation of the Standard Model using the reduction of the gauge boson propagators to local couplings. 
Since the electroweak vertices in Figs.~1 and ~2 are purely leptonic, the reduction to the low-energy, effective Lagrangian coupling is straightforward.  
 
The heavy photon-like W-boson propagator of mass M$_{W}$ can be  
approximated by \cite{sche83} 
\begin{equation} 
\frac{\left\{-q^{\alpha \beta} + (q^{\alpha}_{W} q^{\beta}_{W})/M_{W}^{2}\right\}} 
{[q_{W}^{2} - M_{W}^{2}]} \rightarrow \frac{q^{\alpha \beta}}{M_{W}^{2}} 
\end{equation} 
when the four-momentum transfer q$_{W}$ is such that M$_{W}^{2} >> \mid 
q_{W}^{2}\mid$. The dimensionless electroweak coupling constant g  
of the Weinberg 
theory can be related to the universal Fermi constant G through 
\begin{equation} 
G/\sqrt 2 = g^{2}/(8M_{W}^{2}), 
\end{equation} 
and the effective Lagrangian for neutrino-electron coupling via  
charged electroweak 
currents can be expressed as 
\begin{equation} 
L_{eff}^{c} = {G \over \sqrt 2} 
[\bar{\psi}(\nu)\gamma^{\mu}(1-\gamma_{5})\psi(\nu)][\bar{\psi}(e)\gamma_{\mu}(1-\gamma_{5})\psi(e)]. 
\end{equation} 
The corresponding effective local second-order S-matrix for neutral  
current-induced 
electroweak processes has the form: 
\begin{equation} 
L_{eff}^{n} = {G \over \sqrt 2} 
[\bar{\psi}(\nu)\gamma^{\mu}(1-\gamma_{5})\psi(\nu)][\bar{\psi}(e)\gamma_{\mu}(a-b\gamma_{5})\psi(e)] 
\end{equation} 
where 
 
$a = (-1/2) + 2 sin^{2}\theta_{W}$, $b = (-1/2)$, and  $\theta_{W}$ is the
weak mixing angle\cite{caso,sche83}. 
Channels proceeding via both charged and neutral current corresponding to the $\nu_{e}$-induced $(e^{+}e^{-})$ pairs, can be described by a  
combined effective Lagrangian of the type Eq.(5) given above, 
with 
$a \rightarrow a+1$ and $b \rightarrow b+1$.  $\nu_{\mu}$ -induced $(e^{+}e^{-})$ pair production is described  
by Eq.\ (5). The  V-A limit can be reached for $a \rightarrow 1$ and $b \rightarrow 1$. These limits are given in Table \ref{tab1}.

Using collapsed propagators, the electroweak sector of the diagrams in 
Figs.~1 and ~2 reduce to point coupling, and the  
charged and neutral current channels acquire apparent degeneracy as shown 
in Fig.~3.  
Figure~3 also 
describes the (V-A) limit. We note that the Fierz rotation has been  
applied to the 
charged current channel as customary. 
\subsection{The Nuclear Sector and Its Coupling} 

The nucleus participates via virtual photon exchange with one of the members of the produced flavor-degenerate pair. The virtual photon  
transfers four-momentum q between the lepton propagator and the nucleus, where $q$ is defined through   $q = p-p^\prime$, and $p$ and $p^\prime$ are the initial  
and final four-momenta of the nucleus. Since the nucleus is connected to the electroweak vertex through the virtual photon, the nuclear  
interaction is  purely electromagnetic and only the electromagnetic form factor of the nucleus is operative. As mentioned earlier, we only  
consider spinless nuclei and collisions that leave the nucleus in it's ground 
state.
The nuclear matrix element of the electromagnetic current for a spinless nucleus can be expressed as  
\begin{equation} 
<p'\mid j^{\beta}_{em}(0)\mid p> = (p+p')^{\beta} ~Ze~ F(q^{2}). 
\end{equation} 
The form factor limits the four-momentum transferred to the nucleus to be less than the approximate reciprocal size of the nucleus, $q \le 20$ MeV/c.  
It has been shown by \cite{Sha} that allowing q its full range of allowed kinematical  
domain yields an erroneous E$^{2} ln$ E asymptotic dependence to the cross section instead of the correct E $ln$ E dependence obtained  
using the form factor description. Our studies validate this observation. 
We use an exponential form factor given by \cite{bot} 
\begin{equation} 
F(q^{2}) = \exp\Big(-\mid q^2\mid /\kappa_0^2 \Big), 
\end{equation} 
where, $\kappa_0 = 116.46 A^{\frac{1}{3}} ~ MeV/c$.
 
While we have confined this study to cases where the nuclear response is coherent and elastic, the nucleus in all of these processes could  
respond inelastically. As energies increase, spontaneous release of nucleons or nuclear excitations accompanying the neutrino  
bremsstrahlung by pair radiation may contribute to the total cross sections. We do not consider incoherent contributions from individual  
nucleons in the present work.  At the energies we are interested in (1 to 1000 MeV), these are expected to be smaller than the elastic cross  
sections. 
Contributions from incoherent processes have been studied in earlier work \cite{Cyz,angle}. 

\subsection{The Matrix Elements and Cross Sections} 

The second-order S-matrix elements corresponding to electron-positron pair 
production by neutrinos can be written in a generalized form in terms of the 
appropriate effective Lagrangian as 
\begin{eqnarray} 
M &=& {G\over \sqrt 2} (Ze^{2})F(q^{2}) {P^{\alpha}\over q^{2}} [\bar{U}(k')\gamma^{\mu}(1-\gamma_{5})U(k)] \nonumber \\ & &\bar{U}(r_{- 
})[{\gamma_{\alpha} {1\over (Q{\kern-7pt /}_{-}  -m)} \gamma_{\mu}(a-b\gamma_{5})  + \gamma_{\mu}(a-b\gamma_{5}) {1\over (Q{\kern-7pt /}_{+} -m)}  
\gamma_{\alpha}}]V(r_{+}) 
\end{eqnarray} 
where 
 
$Q_{-} = r_{-} - q$,   $Q_{+} = q - r_{+}$  and $ Q{\kern-7pt /}$ refers to $\gamma_{\alpha} Q^{\alpha}$ in general.
Values of $a$ and $b$ have been defined earlier and differ for the separate cases of  
incident $\nu_{\mu}$ and 
$\nu_{e}$ and for the  
V-A case.  
Equation (8) includes the direct and cross terms. $U$ and $V$ refer to the spinors for the respective particles or antiparticles, and $Q_{-}$ and $Q_{+}$ are the four-momenta of the charged fermion propagators. The four-momentum transfer to the nucleus is $q$, and $e$ is the usual electromagnetic coupling constant. 
$P$ and $q$ are connected to the four-momenta of the initial and final nuclei through  $q = p- p^\prime$ and $P = p+ p^\prime$.  $F(q^{2})$ represents the nuclear  
form factor as discussed earlier and controls the momentum transferred to the nucleus at high energies. It is unity at low energies. The above labeling of the
momenta is shown in Figs.~2 and ~3.  
Introducing the phase space factors, the cross section can be written in 
terms of the matrix element as 
\begin{equation} 
\sigma = {1\over 4E_{\nu}E_{p}} \int {d^{3}k'\over 2E'}{d^{3}r_{+}\over  
2E_{+}}{d^{3}r_{-}\over 
2E_{-}}{d^{3}p'\over 2E_{p'}} \sum_{spins}\mid M\mid ^{2}  
{(2\pi)^{4}\over (2\pi)^{12}} 
\delta^{4}(k'+r_{+} + r_{-} - k-q) 
\end{equation} 
 
Conventions for phase space factors and gamma matrices are those of \cite{bjork}. 
The square of the matrix element is summed over final spins. There is no averaging over initial spins, as the neutrino is helical. The delta  
function ensures four-momentum conservation for the process.

Including flavor variety, the formalism can be readily extended to degenerate 
flavor production beyond the first generation and to nondegenerate  
flavor production using 
appropriate choices for the constants $a$ and $b$.

\section{Standard Model Cross sections} 
The three final leptons and the recoiling nucleus span a twelve dimensional final state phase space. This twelve-dimensional phase space is reduced to eight-dimensions by the delta function constraints. The phase space reduction is carried out in the usual way \cite{phsp}, considering the charged pair in their center of mass frame first and then incorporating the final neutrino. One azimuthal angular  
integration is removed by symmetry arguments and a suitable choice of axis. The remaining seven-dimensional phase space is evaluated  
numerically using Monte Carlo techniques.
 
We display in Fig.~4  a comparison between the earlier results for this cross section obtained by Cyz {\it et. al.} \cite{Cyz} and the charged current results  
for $\nu_{e}$-induced events in nickel. The latter corresponds to the V-A approximation used by \cite{Cyz}  and the nuclear form factor used therein to enable the comparison. For our Standard Model results presented in all other figures, we use the form factor discussed in the previous section.

In reality, the correct cross sections should include neutral current contributions.  Figure~5 displays the cross sections (including neutral currents) for $\nu_{e}$ and $\nu_{\mu}$ induced channels for $^{56}$Fe and $^{208}$Pb. The points represent the values of the numerical computations corresponding to up to 10$^{8}$ Monte Carlo points and a statistical error better than five percent. The $\nu_{\mu}$ induced cross sections are lower than the $\nu_{e}$ initiated ones as expected because the former channels are mediated only by neutral currents, while the latter include both charged and neutral currents.
Clearly, the contributions of both processes increase with $Z$ and are
expected to be most important for high-Z nuclei. 
Iron and lead have been selected as representative nuclei for medium and high Z nuclei that are important in supernova neutrino physics and laboratory experiments, including the terrestrial detection of cosmic neutrinos. A naive $Z^2$ scaling of the results for iron could be obtained to obtain  cross sections on lighter nuclei like oxygen or carbon. 

We have also fitted the energy and $Z^2$ dependence of the 
cross sections so as to represent them by empirical 
expressions that can be used to facilitate applications and estimate the effects of these channels in different environments. The numerical expressions for $\nu_{e}$ and $\nu_{\mu}$ induced channels are given below and were obtained as an empirical fit to the numerically computed cross sections in iron.

The cross sections, in units of cm$^2$,  for first generation pair production by muon- and electron-neutrino bremsstrahlung in iron can be represented as,

\begin{equation}
\ln (\sigma) =  C_{1} x \ln(x)  +  C_{2} x  + C_{3}
\end{equation}

 where $x = \ln E_{\nu} $ and $E_{\nu}$ is the incident neutrino energy, in MeV.

The coefficients $C_{1}, C_2, C_3$ are given in Table \ref{tab2}, and comparisons with numerical calculations are shown in Fig.~5.
The solid lines in Figure 5a correspond to values given by the fitted expressions in Eq.(10) with values from Table \ref{tab2}. A  $Z^{2}$ scaling of the above expression gives excellent fits for both  carbon and lead, demonstrating their usefulness for various applications. The solid curves in Fig. 5b display values obtained using this expression, scaled from iron to lead.

\section{Analytic Reduction} 
 
The complexity of the expression for the spin summed square of the matrix element, with its many terms does not lend itself to analysis of the energy and angular distributions or to the interpretation of the interesting physics of such combined electroweak and photon exchange diagrams. These features however acquire clarity when examined via a reduced form of the matrix element. 
 
The matrix element of Eq.\ (9) can be  
reduced to a compact form by 
using the Dirac equation. Rationalizing the fermion propagators and replacing $Q_{-}$ and $Q_{+}$ by their expressions in terms of $r_{-}$ and $r_{+}$, equation (8) takes the form 
 
 \begin{eqnarray} 
M &=& {G\over \sqrt 2} (Ze^{2})F(q^{2}) {P^{\alpha}\over q^{2}} [\bar{U}(k')\gamma^{\mu}(1-\gamma_{5})U(k)] \nonumber \\ & &\bar{U}(r_{- 
})[{\gamma_{\alpha} {(r_{-}{\kern-5pt /}  - q{\kern-5pt /}  + m)\over (Q_{-}^{2}-m^{2})} \gamma_{\mu}(a-b\gamma_{5})  + \gamma_{\mu}(a-b\gamma_{5})  
{(q{\kern-5pt /}   - r_{+}{\kern-5pt /}  + m) \over (Q_{+}^{2}-m^{2})}  
\gamma_{\alpha}}]V(r_{+}) 
\end{eqnarray} 
 
Rearranging, using usual spinor algebra \cite{bjork}, operating on the electron and positron four momentum vectors from the left and right by their respective spinors, and using the Dirac equation, the matrix element can be expressed as the sum of two terms  
$M = M_{0}+ M^\prime $, where 
\begin{eqnarray} 
M_{0} &=& {G\over \sqrt 2} {(Ze^{2})\over q^{2}}  
[\bar{U}(k')\gamma^{\mu}(1-\gamma_{5})U(k)] \nonumber \\ 
&\times& 2\Big[{(Pr_{-})\over (Q_{-}^{2} - m^{2})} - {(Pr_{+})\over  
(Q_{+}^{2} - m^{2})}\Big][\bar{U}(r_{-})\gamma_{\mu}(a-b\gamma_{5})V(r_{+})] \nonumber \\  
\end{eqnarray} 
and 
\begin{eqnarray} 
M^\prime &=& {G\over \sqrt 2} {(Ze^{2})\over q^{2}}  
\bar{U}(k')[\gamma^{\mu}(1-\gamma_{5})U(k)] \nonumber \\ &\times& \bar{U}(r_{-})  
\big[- {P{\kern-7pt /} q{\kern-5pt /} 
\gamma_{\mu}(a-b\gamma_{5})\over (Q_{-}^{2}-m^{2})} +  
{{\gamma_{\mu} (a-b\gamma_{5}) q{\kern - 5pt /}  P{\kern-7pt /}}  
\over (Q_{+}^{2}-m^{2})} \big] \nonumber \\ &\times&  V(r_{+}). 
\end{eqnarray} 
This decomposition allows us to isolate the part that coincides with the standard electroweak leptonic processes like $\nu -e$ scattering and  
muon decay, for which the spin summations and angular correlations are well known. The nuclear sector with its virtual photon connector is  
also identifiable.  $M_{0}$ allows us to attempt an analytic understanding of the physics of such interfering vertices, which would not otherwise be  
realizable due to the underlying numerical complexity of the terms. The remaining term $M^\prime$ is not amenable to such a compact form. Before assessing the  
contribution of $M^\prime$, we study the behavior of $M_{0}$. 
Neglecting $M^\prime$, the cross section takes the form: 
\begin{equation} 
\sigma = {1\over 4E_{\nu}E_{p}} \int {d^{3}k'\over 2E'}  
{d^{3}r_{+}\over 2E_{+}} {d^{3}r_{-}\over 2E_{-}} 
{d^{3}p'\over 2E_{p'}} \sum_{spins} \mid M_{0}\mid^{2} {(2\pi)^{4}\over  
(2\pi)^{12}} \delta^{4}(k' + r_{+} + 
r_{-} - k - q) 
\end{equation} 
 
The spin summed value of $\mid M_{0}\mid^{2}$ has the form 
\begin{eqnarray} 
\sum_{spins}\mid M_{0} \mid^{2} = {G^{2}\over 2} {(4\pi)^{2}(Z\alpha)^{2}\over 
(q^{2})^{2}} 4 \Big[{(Pr_{-})\over (Q_{-}^{2}-m^{2})} - {(Pr_{+})\over  
(Q_{+}^{2}-m^{2})}\Big]^{2} \nonumber 
\\ 
\times 64[(a+b)^{2}(kr_{+})(k'r_{-}) + (a-b)^{2} (kr_{-})(k'r_{+}) \nonumber \\ + m^{2}(a^{2}-b^{2})(k'k)] 
\end{eqnarray} 
The result of the trace part of this is readily seen to be identical to pure leptonic four-Fermi processes and can be compared by inspection.  
(The sign of the mass term differs from that in neutrino-electron scattering due to the occurrence of a final state positron instead of an incident  
electron.) 
 
We present in Fig.~6, a comparison of the exact numerical calculations with the results obtained using the M$_{0}$ term from the analytic  
reduction. The agreement indicates the neglected terms do not contribute appreciably, and thus the simplified expression of Eq.\ (15) may be  
used for various applications. Expression (15) accounts for about 80 percent of the total cross section.
The connection between the processes studied in this paper and photo-neutrino
pair production can be easily seen by cutting the nuclear legs in the Feynman
diagrams of Fig.s~1--3 \cite{st}. In this case, there is no  
equivalent of the term M$^\prime$, so that one could conclude that in the limit $q^{2} \rightarrow 0$, the term $M^\prime$  should vanish. The fact that usually  
virtual photon processes can be well approximated by photo-nuclear cross sections or the equivalent photon method \cite{bot,Koz,okun} vindicates our  
numerical demonstration of the dominance of $M_{0}$ in $M$. While all cross sections reported here contain the complete matrix element, we have used the reduced ones to compute and explain the distributions reported in the next section.  
 
The survival of the leptonic electroweak sector in factorable form in $M_{0}$ allows us to compare the pair bremsstrahlung with the expressions for neutrino-electron  
scattering and other lowest-order electroweak channels. The angular constraints and distributions typical of weak processes can also be  
recognized through this sector. The factorable nuclear sector imposes the physics constraints typical of virtual photon-exchanged processes,  
and the ultimate distributions and energy behavior are determined by the interference of the two sectors. 
 
\section{The Distributions} 

Angular distributions for all the emitted leptons are shown in Fig.~7
for electron neutrino-lead collisions at 10, 100 and 1000 MeV incident 
energies. The angular distributions are seen to peak in the forward 
direction, the outgoing neutrino having the sharpest rise. All the final leptons display a sharpening of the spectra, emphasizing their forward bias with increasing energy and the angular distribution of the positron is observed to be broader and less forward peaked than that of the electron. The characteristics are discussed below in the light of the conflicting dictates of the electromagnetic and weak vertices.  
  The angular distributions for this process are reported here for the first time and can be used, along with the energy distributions and cross sections to estimate experimental signals.

Energy distributions of the charged leptons are displayed in Figs.~8
and ~9 for lead nuclei. The general nature of these distributions can be explained in terms of the high-energy limit of the analytically reduced matrix elements and can be shown to exhibit the helicity constraints of the electroweak vertex.  
 
Figure~8 provides a comparison of electron and positron energy distributions for incident neutrinos of 10, 100 and 1000 MeV energies. The horizontal axes are scaled by the incident energy. In general, the energy distributions display a differentiation in the spectra of the two members of the produced charged lepton pair, the positron spectra peaking at lower energies and more sharply than that of the electron. The effect is increasingly pronounced with increase of incident and available energy, the positron peak sharpening and shifting 
to lower energies. 

Figure~9 compares the electron and positron distributions for 100 MeV incident neutrinos of muon and electron type respectively with those for the V-A case. It may be noted that the differentiation in the lepton-antilepton spectra displayed in Fig.~8 for the electron neutrino events is retained in the data of Fig.~9, although the effect is more pronounced for the V-A case and least for the muon neutrino channels. The asymmetry between the electron and positron spectra obtained by us is consistent with a similar asymmetry reported for the V-A case by \cite{angle} for second generation pairs produced at higher incident energies. The contributions of the neutral current sectors  
soften the differentiation of the spectra of the two charged leptons for the Standard Model distributions in electron neutrino induced events in contrast to the V-A case. This is why the asymmetry is most pronounced in the V-A limit, corresponding to charged current events only. The muon neutrino produces first generation pairs by neutral current only and the differentiation in the energy distributions of the outgoing pair is barely perceptible. 
 
To understand the distributions let us first look at the high energy limit of the spin summed value of $\mid M_{0}\mid^{2}$, in particular the various four vector products. 
We can write $\mid M_{0}\mid^{2}$ from section V, equation $(15/14)$ as

\begin{eqnarray}
\sum_{spins}\mid M_{0} \mid^{2} = D_{1}  D_{2}  
\end{eqnarray}

where
\begin{eqnarray}
D_{1} & = & {G^{2}\over 2} {(4\pi)^{2}(Z\alpha)^{2}\over
(q^{2})^{2}} 4 \Big[{(Pr_{-})\over (Q_{-}^{2}-m^{2})} - {(Pr_{+})\over 
(Q_{+}^{2}-m^{2})}\Big]^{2} \nonumber
\\
\end{eqnarray}

and
\begin{eqnarray}
D_{2} & = & 64 [ A_{1} (kr_{+})(k'r_{-}) + A_{2}(kr_{-})(k'r_{+})  + A_{3}(k'k)]
\end{eqnarray}

with $A_{1}= (a+b)^{2}$, $A_{2}= (a-b)^{2}$ and $A_{3}= (a^{2}-b^{2}) m^{2}$

The constants a and b have the values stated in earlier sections for specific neutrino flavors, and in Table \ref{tab1}.

We examine the first term, containing four vector products $(kr_{+})$ and $(k'r_{-})$ multiplied by the coefficient $A_{1}$.
In the high-energy limit, for relativistic pairs, the first four-vector product $(kr_{+})$ of this term approaches
$(kr_{+})\rightarrow E_{\nu} E_{+}[1- \cos(\theta_{+}) + O(m^{2})] $ where $\theta_{+}$ represents the angle of the produced positron with respect to the forward direction or the direction of the incident neutrino. This term goes to zero as 
$\cos(\theta_{+})$ goes to unity, and therefore suppresses relativistic positrons in the forward direction. The product reflects the helicity transfer at the weak vertex and originates in the charged current component of the combined process. The behaviour is reminiscent of similar helicity constraints that influence neutrino- and antineutrino- electron scattering at high energies and distributions
in muon and pion decay \cite{okun}.

The suppression of relativistic forward positrons, dictated by the weak vertex is opposed by the virtual photon sector that links the weak vertex to the recoiling nucleus. The presence of $q^2$ in the denominator of $D_{1}$ yields the largest probability for those events with the smallest values of $q^2$. This tends to focus all final leptons into a narrow forward cone at high energies,  since energy-momentum conservation takes the form:  $ k \approx k^{\prime} + r_{+} + r_{-}$, 
for small $q^2$.
This is common to many other processes at high energies, including neutrino-electron scattering\cite{okun} and pair production\cite{bot}. As $q^2$ occurs as a square in the denominator it is powerful in maintaining the clustering of all final leptons into a narrow forward cone, with an opening angle of the order of $(m/E_{\nu})^2$. 
The analytic expression for the spin-summed square of the matrix element allows us to understand the competing effects that influence the final distribution.  The conflict between the forward cone dictates of usual high-energy exit channel behavior and the strong suppression of forward positrons by the weak vertex and its helicity constraints can be best compromised by suppressing high energy positrons and enhancing positron population of lower energy states. This behaviour is similar to the suppression of pion decay by the electron mode despite its favoring by phase space as compared to the dominant muon mode.  
 
The above arguments are supported by the angular and energy and  distributions in Figs.~7,8, and 9. All final leptons dominantly go forward in accordance with traditional pair production observations, and this trend is enhanced at higher energies. On the other hand, to satisfy the helicity requirements of the charged current component, the positron energy distribution peaks at lower energies and is strongly suppressed at high energies. This is the direct effect of the term $kr_{+}$ which goes to zero for relativistic forward positrons. As incident energies increase, the charged leptons would normally tend to become  increasingly relativistic and the differentiation in their spectra due to suppression of relativistic forward positrons is accentuated further as seen in 
Fig.~7. This behavior is also consistent with 
helicity arguments similar to those that explain the asymmetry in neutrino-antineutrino scattering at high energy, correlations in muon decay and other weak interaction processes.  The second terms do not display this requirement and their interference softens the asymmetry in the distributions as compared to 
the V-A case.

Turning to the electron spectra, the second four vector product $(k'r_{-})$, involving the electron and the final neutrino, reflects the helicity requirements on these leptons and a similar suppression on their collinearity. However, this is different from the previous case, where the incident neutrino defines the forward direction. For the electron-final neutrino correlation term, both electron and final neutrino have freedom of direction. This allows a spread to the correlation between them. However, this too suppresses forward electrons, collinear with the outgoing strongly forward neutrino, and hence peaks the electrons at intermediate energies. The forward suppression is milder for the electrons as it occurs relative to the final neutrino. This differs from the case of the positron, which is suppressed with respect to colinearity with the incident neutrino which in fact defines the forward direction. This can explain the sharper and lower energy peak for the positron as compared to the electron and the stronger forward production of electrons.  

We now investigate the importance of the first term 
that favors lower positron energies compared to the others in $D_{2}$ 
and through it on the expression for $\sigma$ by quantitative comparisons 
of the coefficients. Table \ref{tab3} displays the numerical values of the coefficients $A_{1}$ and $A_{2}$ in the expression for  $D_{2}$  for the V-A,  $\nu_{e}$ and $\nu_{\mu}$ cases. (The interference term, characterized by the coefficient $A_{3}$ is mass dependent and does not contribute much at high energies). 

It is apparent from Table \ref{tab3}, that $A_{1}$ dominates for $\nu_{e}$ and pure charged current (or V-A) events, the dominance being more for the V-A case when the second coefficient is zero. This explains the strong differentiation in the charged lepton spectra for these classes of events. The presence of the second term, arising from the interference from the neutral current contributions, softens the $\nu_{e}$ spectra slightly, compared to the V-A case. On the other hand, for $\nu_{\mu}$ events, both coefficients are of similar magnitude, giving rise to the much softer differentiation in the charged lepton spectra for these events.

The energy and angular distributions reported here have been calculated in the laboratory frame using the reduced matrix elements 
to enable comparison with corresponding analytic expressions. A check with the complete matrix elements, indicated that the agreement between the reduced and complete matrix elements is even better for the distributions than for the cross sections. Further, the distributions are of use and interest mainly to display the underlying physics and to be indicative of experimental signals. This contrasts our interest in the cross-sections, which are useful to obtain numerical estimates of the influence of these processes on neutrino physics in general. We reiterate that cross section results in Figs.~4-6 include the full matrix element as do the empirical 
fits that represent their energy dependence.

\section{Experimental Signatures and Possible Roles\\ in Neutrino Physics}

From the laboratory perspective, the event signature would be the 
produced pair, both members being emitted predominantly in the forward direction. The opening angle of the outgoing positron is larger than that of the electron as discussed earlier. Further, the positron energies peak at the lower energy end of the spectrum while electron energies have a broader distribution of energies. Events of this type could be studied by detecting both components of the produced pair in coincidence. Such
pairs should be distinct from those arising from de-excitation of 
nuclei, as the energy
available to the latter pair is fixed by the de-excitation energy of 
the nucleus and the
correlation signature of such a de-excitation pair.  In contrast, the energy
distributions of the neutrino-initiated direct pairs display 
continuum characteristics, and
the angular distributions respect helicity constraints of the 
electroweak vertex and the
physics of the photon vertex.

If the primary experimental interest is not to detect the bremsstrahlung process itself, but rather its influence on measurements of the lowest order neutrino-nucleus events or neutrino flux determinations, this can be estimated using the theoretical cross sections presented in this paper and the empirical formula obtained for them. A major focus of current neutrino science is the detection of neutrinos of extra-terrestrial origin, - those from the sun or from supernovae. The primary event signatures for these are from charged current neutrino interactions occurring on the detector target nuclei.
Detection processes that measure the outgoing electron in neutrino induced charged current events could pick up pseudo-events generated by electrons from bremsstrahlung events, particularly in high Z targets like lead or iodine. As experimental constraints tighten and increased accuracy in event rates is attempted, it becomes important to estimate the effects of backgrounds such as the pair bremsstrahlung investigated in this paper. As mentioned earlier, the pair bremsstrahlung cross sections of order $(GZ\alpha)^2$ may become comparable to order alpha corrections to the lowest order electroweak channels and $G^2 \alpha$ order processes for values of Z exceeding 12.

The process could also be a potential source of background to neutrino-electron scattering experiments and experiments that use $\nu -e$ scattering to establish flux measurements and normalizations. This background will be increasingly important for higher Z values of the nuclei of the material hosting the participating electrons. Even for light nuclei, the process could be a source of background for experiments that seek flux-accuracy to a few percent, or better.

We present pair bremsstrahlung cross sections folded with neutrino energy spectra for neutrinos emanating from thermal sources with typical supernovae temperatures in Table \ref{tab4}. We compare our results for iron and lead with those for neutrino-electron scattering. We have used for the neutrino distributions, the normalized Fermi-Dirac form given in \cite{hax} for appropriate supernova temperatures.  As expected, the contribution of pair bremsstrahlung increases as a relative percentage of the neutrino-electron scattering for increasing Z and energy values. 

The results in Table \ref{tab4} include thermal sources at temperatures higher than those expected to generate the bulk of electron neutrinos, which decouple at temperatures lower than the higher flavor neutrinos. This inclusion allows for their use in applications involving supernova thermometry that explore oscillations from tau and muon to electron flavors \cite{hax,kl}. Muon or tau neutrinos that have oscillated to electron flavor would interact in a terrestrial detector according to the electron neutrino values at the appropriately higher temperatures. 

A comment about anti-neutrino events is in order. The diagrams of figures~1--3
can be easily modified to describe anti-neutrino induced events. As a 
particle-antiparticle pair is produced, crossing symmetry dictates identical cross sections for incident neutrinos or anti neutrinos. This has been discussed in additional detail in \cite{Cyz}. However, the distributions in 
figures~7--9 would display an interchange between the charged leptons due to the well-known flipping of the charged current part (appearing through the first term in $D_{2}$) as neutrino-electron scattering changes from incident neutrinos to antineutrinos. This does not alter the final cross sections and the results presented here for neutrinos would apply equally to antineutrinos. 

Apart from the influence of pair bremsstrahlung on the detection of cosmic neutrinos and hence on our understanding of the astrophysical events they portray, the question of possible roles in the astrophysical dynamics themselves also arises.  What 
might be the astrophysical
environments to spawn such neutrino-bremsstrahlung pairs?  High Z 
environments such as
pre-collapse supernova cores, scenes of the stalled shock, or the 
neutrino-driven winds hosting
the r-process and neutrino nucleosynthesis suggest themselves as 
potential sites.  The density
of the Fermi sea, however, would inhibit low-energy electron 
production in pre-collapse cores. It may be noted that the Coulomb distortion effect would affect low-energy electron and positron emission oppositely and could be 
considered, along with the Fermi sea effects, to determine the cross sections in physical supernova environments. 

The question does arise whether the pair bremsstrahlung might be more important for anti neutrinos in supernova environments with a high population of neutron-rich nuclei. As the lowest order nuclear channels are suppressed due to non- availability of neutron states as well as the negative effects of the Coulomb barrier \cite{kl}, other channels such as pair bremsstrahlung may provide alternate means of slowing anti-neutrinos.
 
\newpage 
\noindent 
 
\section*{Acknowledgments} 

We would like to thank Professor Leo Stodosky for 
pointing out the possibility of using the neutrino-nucleus pair  
production as an 
experimental tool to study the photon-neutrino cross sections.\cite{st} 
 
This research was sponsored by the Division of Nuclear Physics of the  
U.\ S.\ Department of 
Energy under Contract No.\ DE-AC05-00OR22725 managed by UT-Battelle, 
LLC, and in part by the Laboratory Directed Research and Development 
Program of Oak Ridge National Laboratory, managed by UT-Battelle, 
LLC, for the U. S. Department of Energy under Contract No. DE-AC05-00OR22725.
Lali Chatterjee would like to acknowledge research support by the National 
Science Foundation Grant No. PHY-0074759 and Jianshi Wu would like to 
acknowledge DOE Grant No. DE-FG02-97ER41044 and the North Carolina Supercomputing Center for their support.

\begin{table}
\begin{center}
\caption{ Given in the table are the dimensionless parameters, a,b, 
of the effective Lagrangian for different flavors of the incident neutrino. }
\vspace{0.3cm} 
\begin{tabular}{|c|c|c|}
\raisebox{0pt}[12pt][6pt]{Neutrino Flavor~}  & 
\raisebox{0pt}[12pt][6pt] {a~} &
\raisebox{0pt}[12pt][6pt] {b~} \\
\hline
\raisebox{0pt}[12pt][6pt]{$\nu_e$~}  & 
\raisebox{0pt}[12pt][6pt] {$2 \sin^2(\Theta_W)-\frac{1}{2}$~}   &
\raisebox{0pt}[12pt][6pt]{$-\frac{1}{2}$ ~} \\ 
\hline
\raisebox{0pt}[12pt][6pt]{$\nu_\mu$~}  & 
\raisebox{0pt}[12pt][6pt] {$2 \sin^2(\Theta_W)+\frac{1}{2}$~}   &
\raisebox{0pt}[12pt][6pt]{$+\frac{1}{2}$ ~} \\ 
\hline
\raisebox{0pt}[12pt][6pt]{$V-A$~}  & 
\raisebox{0pt}[12pt][6pt] {$+1$~}   &
\raisebox{0pt}[12pt][6pt]{$+1$ ~} \\ 
\end{tabular}
\label{tab1}
\end{center}
\end{table}
\begin{table}
\begin{center}
\caption{ Values of the coefficients $C_1,C_2,C_3$ used in Eq.(10) for 
representing the total pair Bremsstrahlung cross sections for both 
electron and muon incident neutrinos on iron.}
\vspace{0.3cm} 
\begin{tabular}{|c|c|c|c|}
\raisebox{0pt}[12pt][6pt]{Neutrino Flavor~}  & 
\raisebox{0pt}[12pt][6pt] {$C_1$~} &
\raisebox{0pt}[12pt][6pt] {$C_2$~} &
\raisebox{0pt}[12pt][6pt] {$C_3$~} \\
\hline
\raisebox{0pt}[12pt][6pt] {$\nu_e$~}  & 
\raisebox{0pt}[12pt][6pt] {$0.34032$~}   &
\raisebox{0pt}[12pt][6pt] {$5.15593$~}   &
\raisebox{0pt}[12pt][6pt] {$109.5118$ ~} \\ 
\hline
\raisebox{0pt}[12pt][6pt]{$\nu_\mu$~}  & 
\raisebox{0pt}[12pt][6pt] {$0.37696$~}   &
\raisebox{0pt}[12pt][6pt] {$4.64135$~}   &
\raisebox{0pt}[12pt][6pt] {$110.4906$ ~} \\ 
\end{tabular}
\label{tab2}
\end{center}
\end{table}
\begin{table}
\begin{center}
\caption{ Given in the table are the dimensionless parameters, $A_1$, and 
$A_2$ of (Eq.s (16) -- (18))for different flavors of the incident neutrino. }
\vspace{0.3cm} 
\begin{tabular}{|c|c|c|}
\raisebox{0pt}[12pt][6pt]{Neutrino Flavor~}  & 
\raisebox{0pt}[12pt][6pt] {$A_1$~} &
\raisebox{0pt}[12pt][6pt] {$A_2$~} \\
\hline
\raisebox{0pt}[12pt][6pt]{$\nu_e$~}  & 
\raisebox{0pt}[12pt][6pt] {$2.13$~}   &
\raisebox{0pt}[12pt][6pt]{$0.21$ ~} \\ 
\hline
\raisebox{0pt}[12pt][6pt]{$\nu_\mu$~}  & 
\raisebox{0pt}[12pt][6pt] {$0.29$~}   &
\raisebox{0pt}[12pt][6pt]{$0.21$ ~} \\ 
\hline
\raisebox{0pt}[12pt][6pt]{$V-A$~}  & 
\raisebox{0pt}[12pt][6pt] {$4$~}   &
\raisebox{0pt}[12pt][6pt]{$0$ ~} \\ 
\end{tabular}
\label{tab3}
\end{center}
\end{table}
\begin{table}
\begin{center}
\caption{ Neutrino induced pair cross sections off of iron and lead
for normalized Fermi--Dirac spectra of temperature $T$ are compared to
those for $(\nu_e,e^-)$ elastic scattering. 
}
\vspace{0.3cm} 
\begin{tabular}{|c|c|c|c|}
\raisebox{0pt}[12pt][6pt]{T (MeV)}  & 
\raisebox{0pt}[12pt][6pt] {$\sigma_e \times 10^{-42} cm^2$~} &
\raisebox{0pt}[12pt][6pt] {$\sigma_{pair}(Fe)/\sigma_e$~} &
\raisebox{0pt}[12pt][6pt] {$\sigma_{pair}(Pb)/\sigma_e$~} \\
\hline
\raisebox{0pt}[12pt][6pt]{$4.0$~}  & 
\raisebox{0pt}[12pt][6pt] {$0.113$~}   &
\raisebox{0pt}[12pt][6pt]{$0.007$ ~}  &
\raisebox{0pt}[12pt][6pt]{$0.103$ ~}   \\
\hline
\raisebox{0pt}[12pt][6pt]{$6.0$~}  & 
\raisebox{0pt}[12pt][6pt] {$0.170$~}   &
\raisebox{0pt}[12pt][6pt]{$0.018$ ~}  &
\raisebox{0pt}[12pt][6pt]{$0.180$ ~}  \\
\hline
\raisebox{0pt}[12pt][6pt]{$8.0$~}  & 
\raisebox{0pt}[12pt][6pt] {$0.323$~}   &
\raisebox{0pt}[12pt][6pt]{$0.026$ ~}  &
\raisebox{0pt}[12pt][6pt]{$0.258$ ~}  \\
\hline
\raisebox{0pt}[12pt][6pt]{$10.0$~}  & 
\raisebox{0pt}[12pt][6pt] {$0.328$~}   &
\raisebox{0pt}[12pt][6pt]{$0.034$ ~}  &
\raisebox{0pt}[12pt][6pt]{$0.335$ ~}  \\
\hline
\raisebox{0pt}[12pt][6pt]{$12.0$~}  & 
\raisebox{0pt}[12pt][6pt] {$0.342$~}   &
\raisebox{0pt}[12pt][6pt]{$0.041$ ~}  &
\raisebox{0pt}[12pt][6pt]{$0.412$ ~}  \\
\hline
\raisebox{0pt}[12pt][6pt]{$15.0$~}  & 
\raisebox{0pt}[12pt][6pt] {$0.425$~}   &
\raisebox{0pt}[12pt][6pt]{$0.053$ ~}  &
\raisebox{0pt}[12pt][6pt]{$0.525$ ~}  \\
\end{tabular}
\label{tab4}
\end{center}
\end{table}
\newpage 
\begin{figure} 
\begin{center} 
\epsfysize=6.0in  
\epsfbox{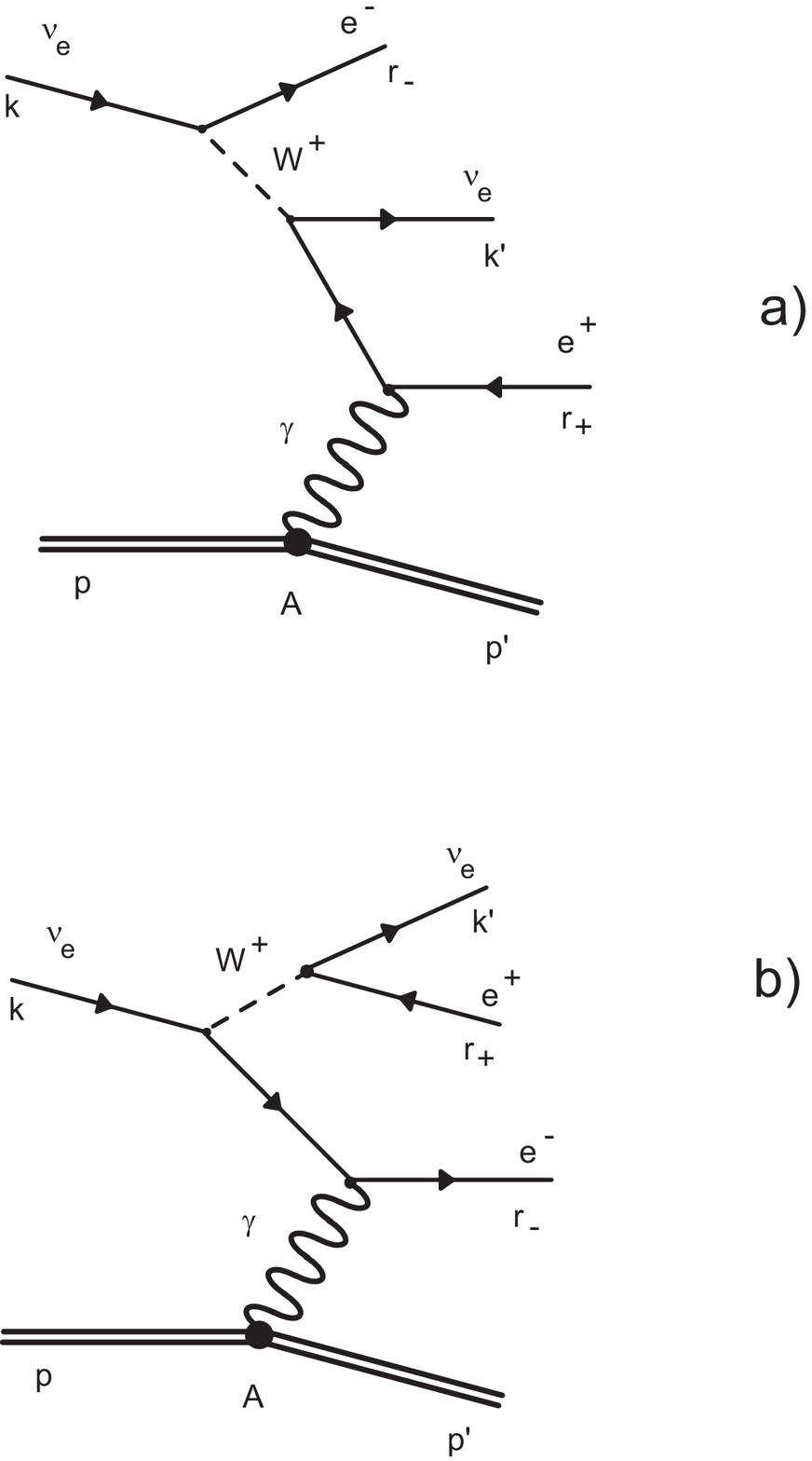}  
\caption{ Feynman diagrams for charged current mediated channels, a) direct, 
and b) cross diagrams for the virtual photon exchange.}
\end{center} 
\label{fig1} 
\end{figure} 
%
\begin{figure} 
\begin{center} 
\epsfysize=3.0in  
\epsfbox{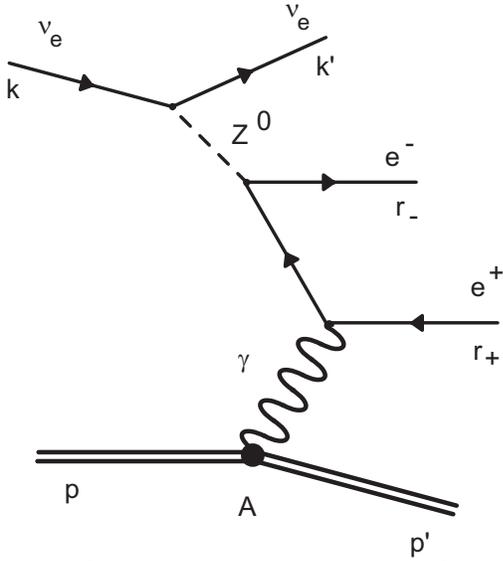}  
\caption{ Feynman diagram for neutral current mediated channels. The cross 
diagram is obtained by $e^{+} \leftrightarrow e^{-}$.}
\end{center} 
\label{fig2} 
\end{figure} 
%
%
\begin{figure} 
\begin{center} 
\epsfysize=3.0in  
\epsfbox{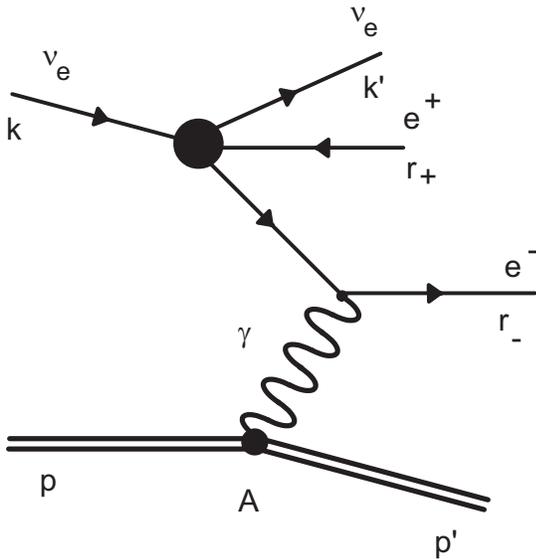}  
\caption{Feynman diagram for collapsed gauge boson propagators. The cross diagram is 
obtained by $e^{+} \leftrightarrow e^{-}$.}
\end{center} 
\label{fig3} 
\end{figure} 
\newpage 
\begin{figure} 
\begin{center} 
\epsfysize=8.0in  
\epsfbox{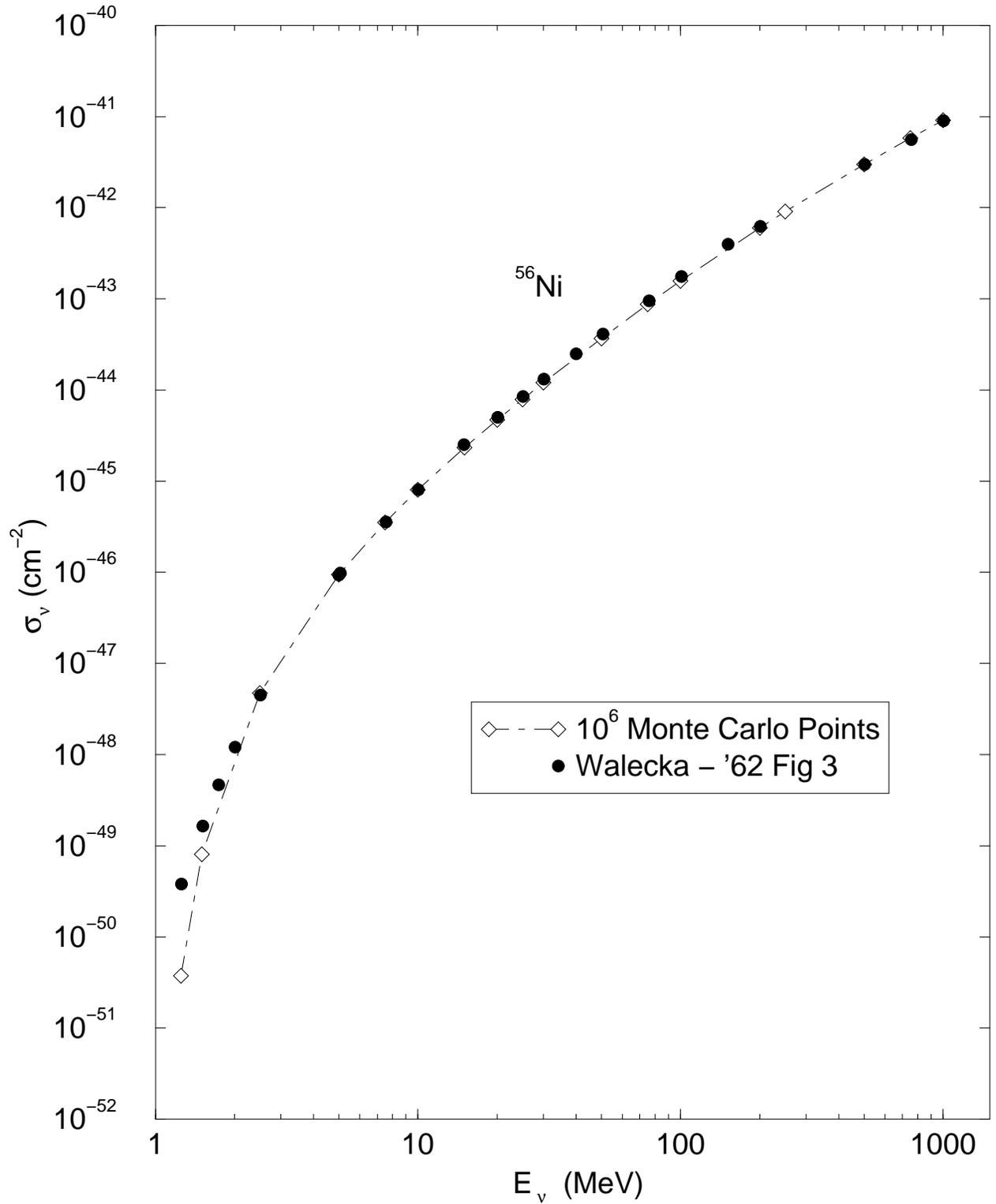}  
\caption{Comparison of neutrino induced pair cross sections, as a function 
of laboratory neutrio energy,  off of  nickel using the V--A approximation with the earlier results of Ref.~15.} 
\end{center} 
\label{fig4} 
\end{figure} 
\newpage 
\begin{figure} 
\begin{center} 
\epsfysize=8.0in  
\epsfbox{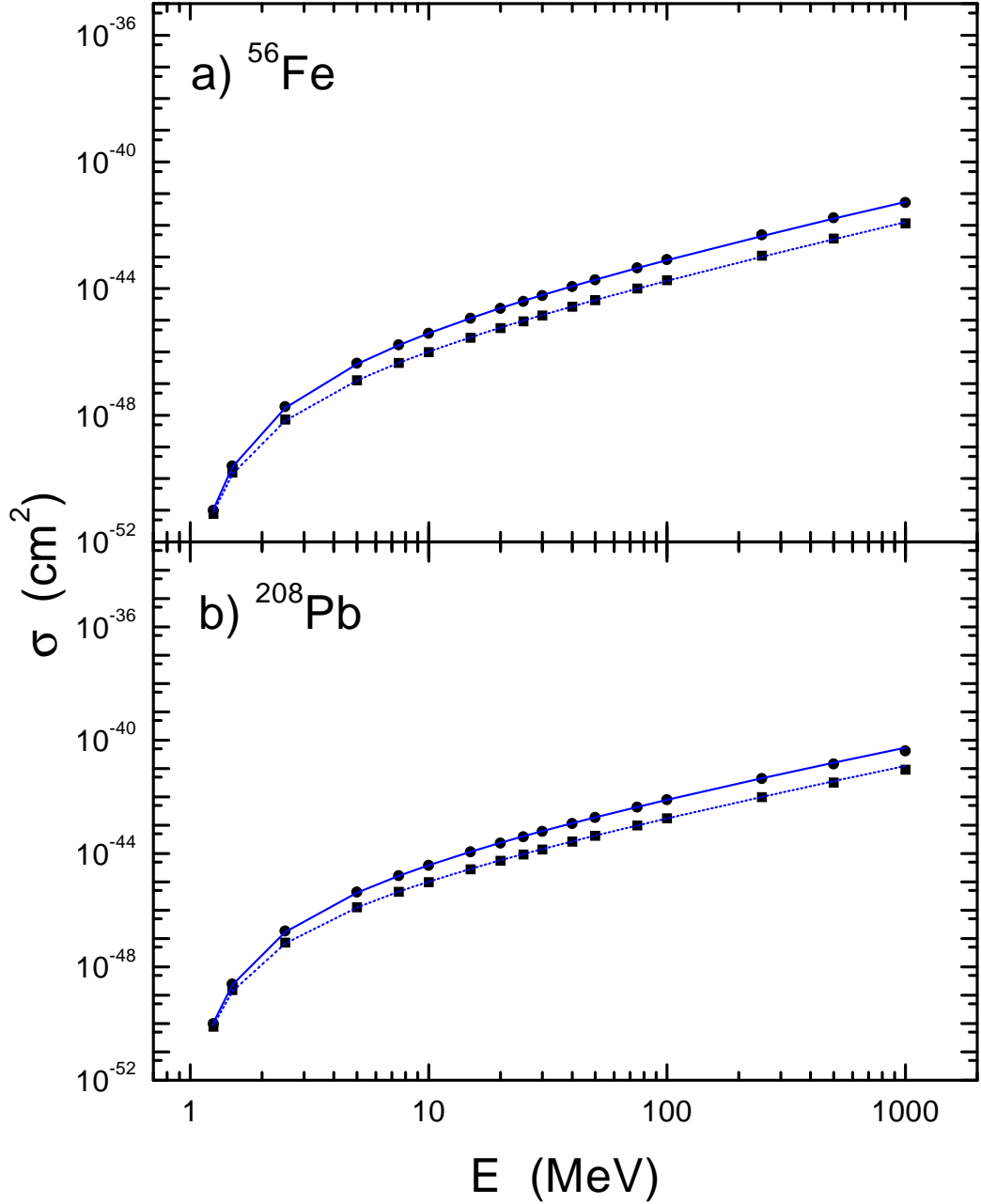}  
\caption{a) Electron-neutrino (solid circles, and solid line), and 
muon-neutrino (solid squares and dashed line), initiated Standard Model 
pair cross  sections on iron evaluated numerically (circles and squares) and 
compared to the empirical cross section formulas, Eq.(10) (solid and 
dashed lines),  b) same as in a) for the nucleus lead, for this case, 
the solid and dashed lines represent Eq.(10) with an assumed $Z^2$ scaling.} 
\end{center} 
\label{fig5} 
\end{figure} 
\newpage 
\begin{figure} 
\begin{center} 
\epsfysize=8.0in  
\epsfbox{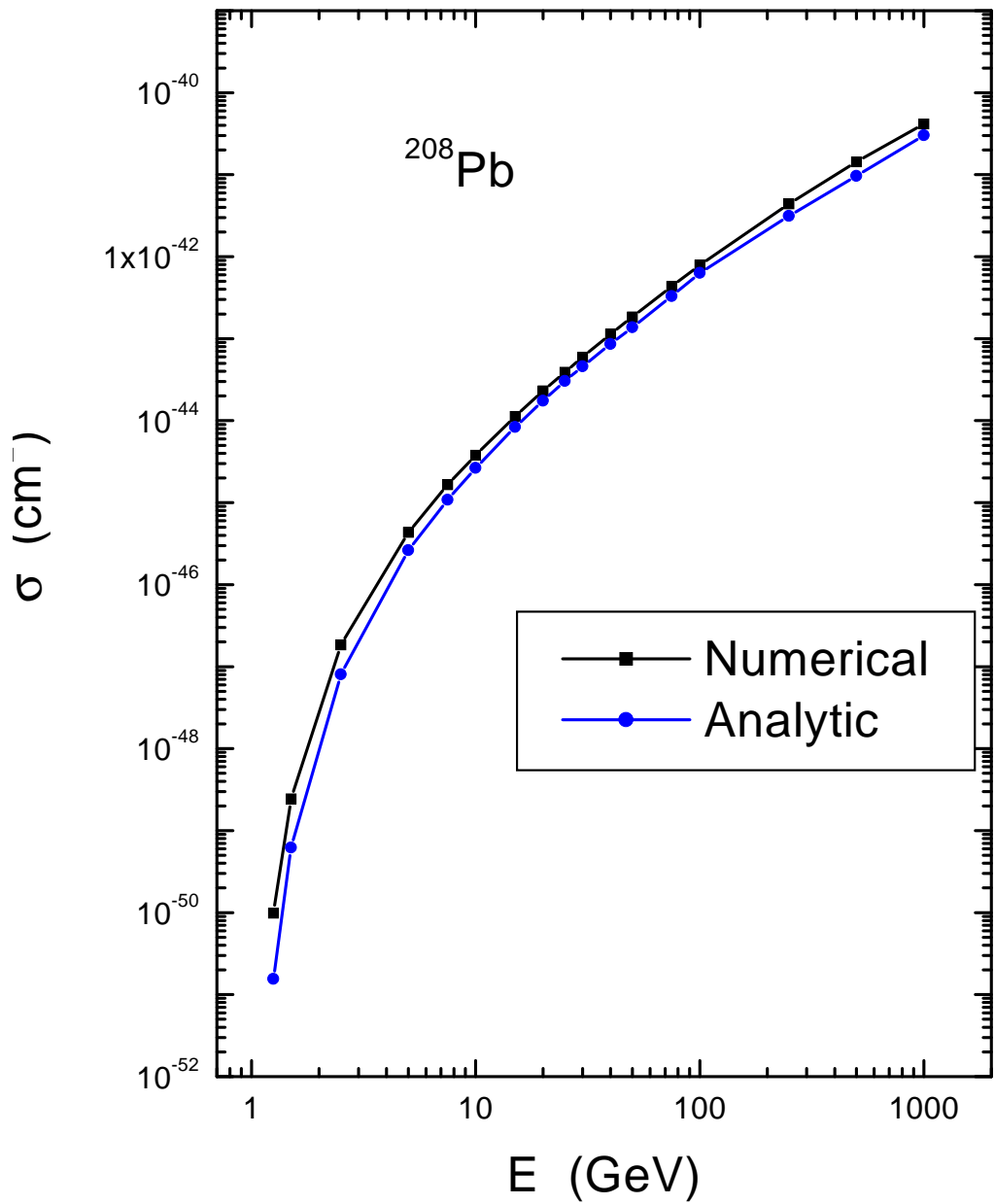}  
\caption{ Electron-neutrino induced pair cross sections off of lead as a 
function of the laboratory neutrino energy, computed numerically, and 
compared those obtained using the analytic approximation discussed in the 
text} 
\end{center} 
\label{fig6} 
\end{figure} 
\newpage 
\begin{figure} 
\begin{center} 
\epsfysize=8.0in  
\epsfbox{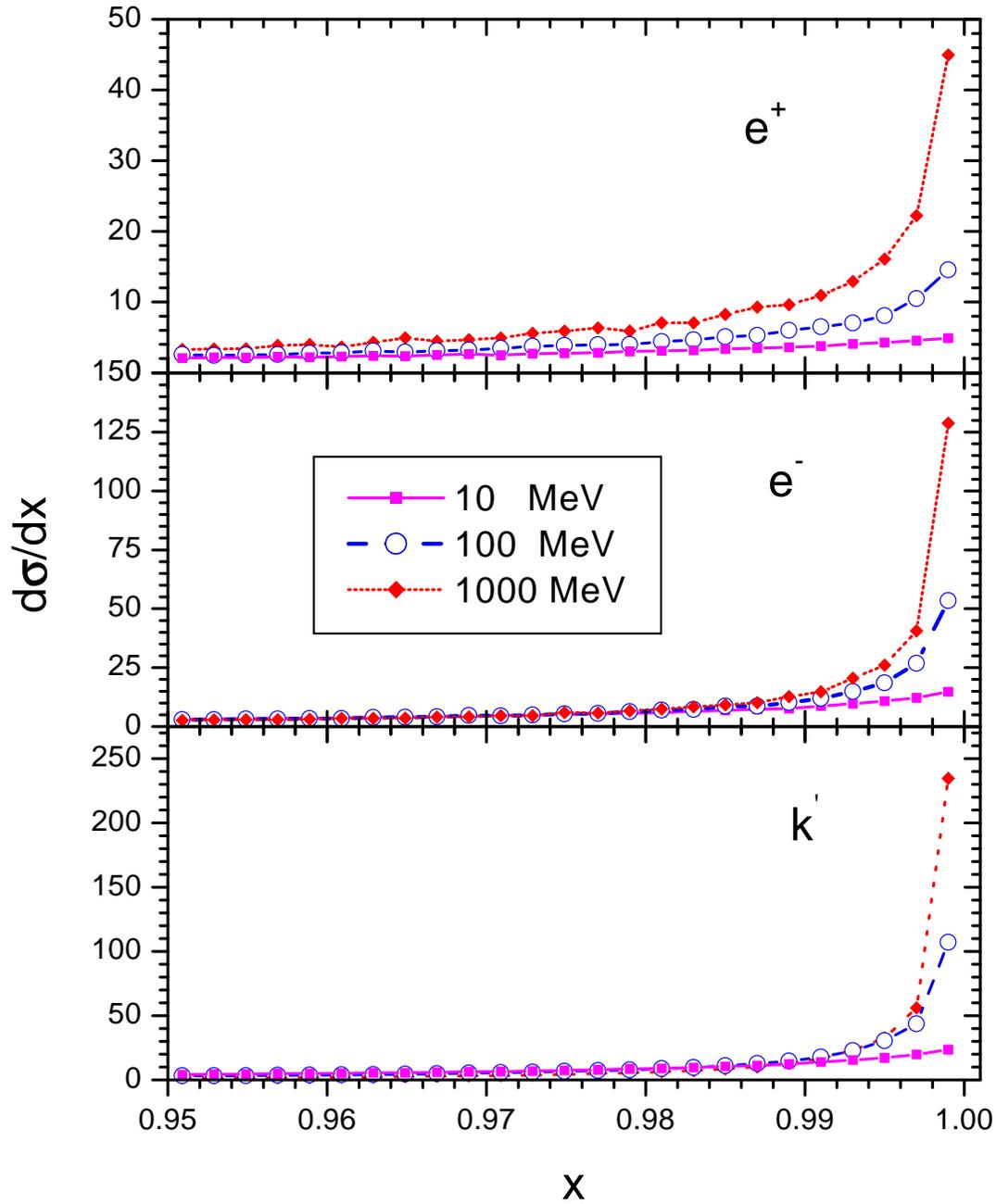}  
\caption{Distributions for the electron, positron and the outgoing 
neutrino, as a function of $x$, the cosine of the outgoing angle with 
respect to the ingoing neutrino direction,  for electron neutrinos of 
incident energies  10, 100, and 1000  
MeV on lead. Note that all three distributions are sharply peaked in the
forward direction, the positron distribution being less so than the others.
The distributions are normalized so that $\int dx d\sigma/dx = 1$.} 
\end{center} 
\label{fig7} 
\end{figure} 
\newpage 
\begin{figure} 
\begin{center} 
\epsfysize=7.0in  
\epsfbox{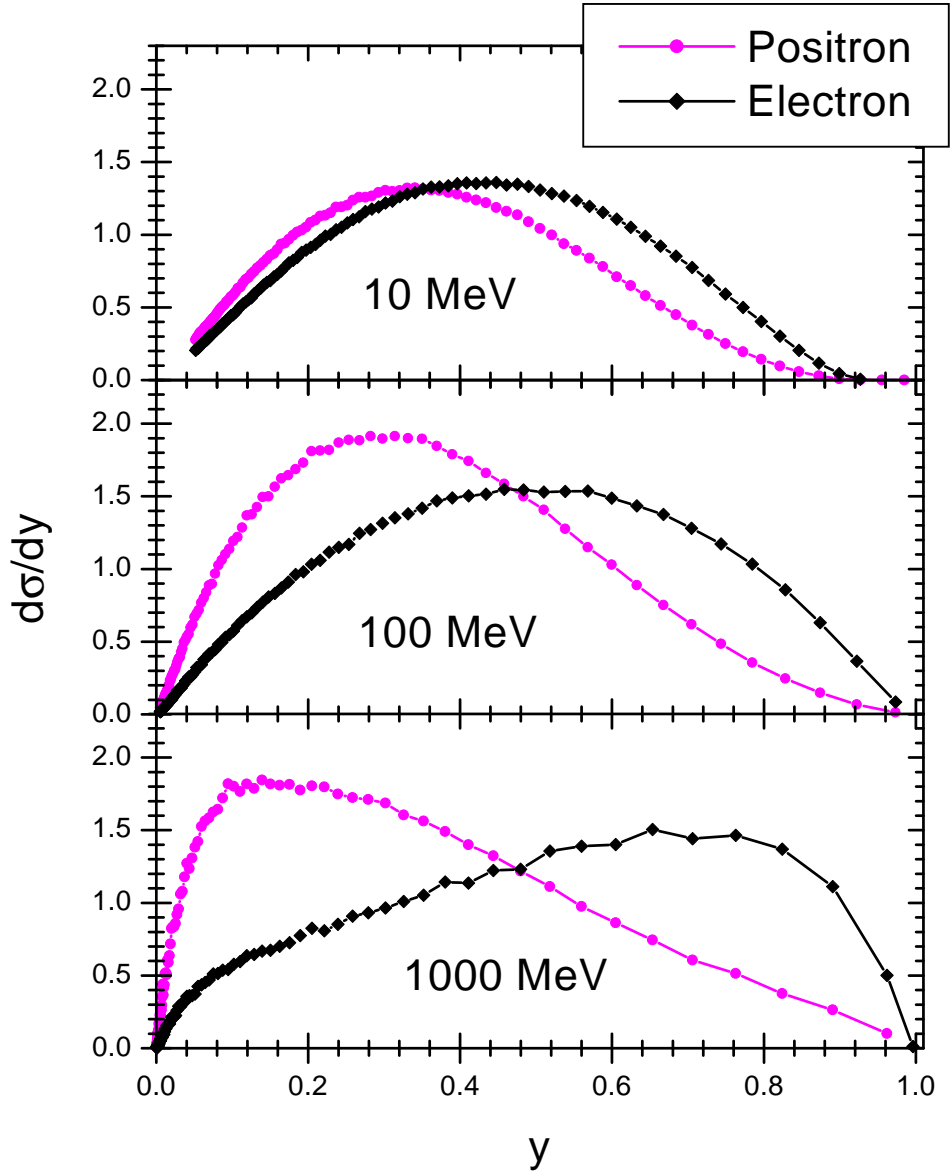}  
\caption{ Energy distributions for electrons and positrons for  
electron neutrinos incident on lead with 
energies of 10, 100, and 1000 MeV.
The distributions are normalized so that $\int dy d\sigma/dy = 1$,
$y = E/E_\nu$, where $E_\nu$ is the energy of the ingoing neutrino.} 
\end{center} 
\label{fig8} 
\end{figure} 
\newpage 
\begin{figure} 
\begin{center} 
\epsfysize=8.0in  
\epsfbox{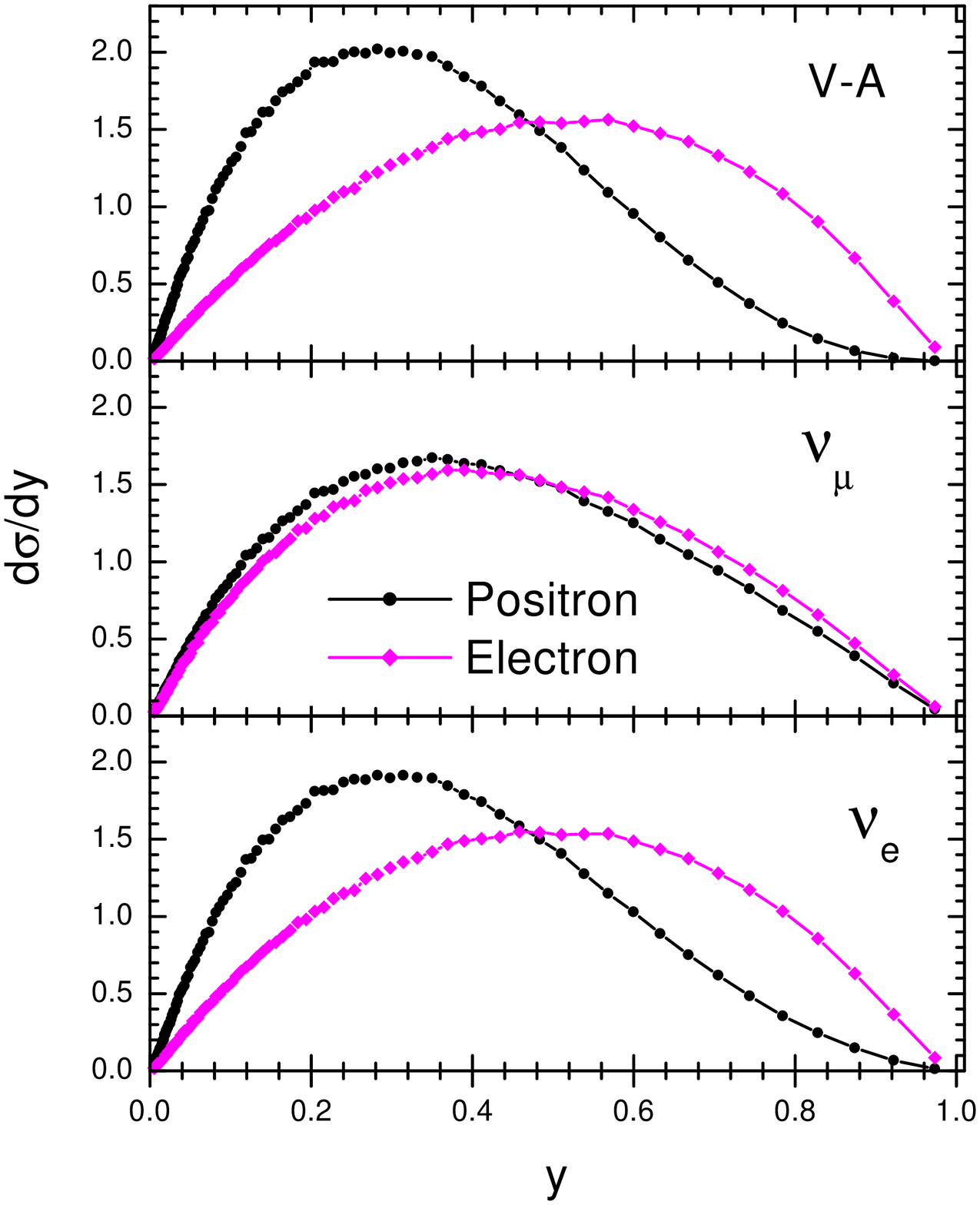}  
\caption{ Energy distributions for electrons and positrons for neutrino 
incident on lead at an energy of 100 MeV  
for electron neutrinos, muon neutrinos, and for the case (V-A).
The distributions are normalized so that $\int dy d\sigma/dy = 1$,
$y = E/E_\nu$, where $E_\nu$ is the energy of the ingoing neutrino.} 
 
\end{center} 
\label{fig9} 
\end{figure} 

\begin{thebibliography}{unsrt} 

\bibitem{kltv} E. Kolbe, K. Langanke, F. K. Thielemann and P. Vogel, \Journal{\PRC}{52}{3437}{1995}.

\bibitem{avi} F. T. Avigone III and Yu. V. Efremenko, \Journal{Nucl. Phys. B (Proc. Suppl.)}{87}{304}{2000}.

\bibitem{lsw} L.Chatterjee, M.R.Strayer and J.S. Wu, {Proc. Of the Carolina Symposium on Neutrino Physics, March 10-12, Columia, SC. Ed. K. Kubodera, World Scientific (in press)}.

\bibitem{schin}P. J. Schinder {\it et al}, {\em Astrophysical Journal}  
{\bf 313}, 531 (1987). 

\bibitem{dic72} D.A. Dicus, \Journal{\PRD}{6}{941}{1972}. 

\bibitem{dic97}D. A. Dicus and W. W. Repko, \Journal{\PRL}{79}{569}{1997}. 

\bibitem{bern} J. Bernabeu, S. M. Bilenki, F. J. Botella and J. Segura, \Journal{Nucl. Phys. B}{246}{434}{1994}.

\bibitem{lee} T. D. Lee and A. Sirlin, \Journal{Rev. Mod. Phys}{36}{666}{1964}.
 
\bibitem{bahc95}J. N. Bahcall, M. Kamionkowski, and A. Sirlin,  
\Journal{\PRD}{51}{6146}{1995}. 

\bibitem{sche83}F. Scheck in {\em Leptons, Hadrons and Nuclei} (North  
 Holland Publishing, 301, 1983). 

\bibitem{kl} E. Kolbe and K. Langanke, \Journal{\PRC}{63}{025802-1}{2001}.

\bibitem{hax} W. C. Haxton, \Journal{\PRD}{36}{2283}{1987}.

\bibitem{woo90}S. E. Wooseley {\it et al}, {\em Astrophysical  
Journal} {\bf 356}, 272 (1990). 

\bibitem{con72} J. S. O'Connell,  
T. W. Donnelly, and J. 
D. Walecka, {\em Phys.\ Rev.}\ C {\bf 6}, 719 (1972). 

\bibitem{Cyz}W. Czyz, G. C. Sheppey, and J. D. Walecka, {\em Nuovo  
Cimento} {\bf 34}, 404 
(1964). 

\bibitem{Koz}M. A. Kozhushner and E. P. Shabalin, {\em Sov.\ Phys.\ JETP}  
{\bf 14}, 676 (1962). 
\bibitem{Sha} E. P. 
Shabalin, {\em Sov.\ Phys.\ JETP} {\bf 16}, 125 (1963). 

\bibitem{Mar} M. S. Marinov, Yu. P. Nikitin, Yu. P. Orevkov and E. P. Shabalin, \Journal{Sov. Jour. Of Nucl. Phys} {3}{678}{1966}

\bibitem{okun}L. B. Okun in {\em Leptons and Quarks} (North-Holland  
Publishing Co., Amsterdam, 
1982);  

\bibitem{angle}J. Lovseth and M. Radomski, Phys. Rev. D3, 2686 (1971) 
 
\bibitem{caso}C. Caso {\it et al}, {\em European Physical Journal} C  
{\bf 3}, 1 (1998). 
 
\bibitem{bot}C. Bottcher and M. R. Strayer,  
\Journal{\PRD}{29}{1330}{1989}; M. Dress, 
J. Ellis, and D. Zeppenfeld, \Journal{\PLB}{223}{454}{1989}. 
 
 
\bibitem{bjork}J. D. Bjorken and S. D. Drell in {\em Relativistic  
Quantum Mechanics} (McGraw-Hill 
Book Co., 1964). 

\bibitem{phsp} V. D. Barger and Roger J. N. Phillips,in {\em 
Collider Physics} (Addison-Wesley Publishing Co., 1987) 

\bibitem{st}L. Stodosky (Munich), private communication, March 12, 2000;
J. Physique, {\bf 35}, (1974)
{\it Intl. Colloquium on Photon-Photon Collsion}
Coulomb and Photon Effects at High Energy.
\end{thebibliography}
\end{document}